\documentclass[
  , draft            
  ]
  {aipproc}

\layoutstyle{6x9}

\begin{document}

\title{Accessing the Gluon Polarization through $A_{LL}$ in $\pi^{\pm}$ Production at PHENIX.}

 


\author{Astrid Morreale on behalf of the PHENIX collaboration}{
  address={University of California, Riverside, CA, 92521 USA}
}

\begin{abstract}
   With proton-proton collisions at PHENIX a variety of direct channels are used to probe the proton substructure. Of the many channels available at PHENIX, charged pion measurements are expected to have sensitivity of $\Delta$g and thus help in the global analysis that will constrain it. We present a measurement of mid-rapidity 
  charged pion production double longitudinal spin asymmetries ($A_{LL}$) at the $p_{T}$ range of 5-10 GeV/c at collision energies of $\sqrt{s}=200$ GeV.  
\end{abstract}
\maketitle
\section{Introduction}
    Measurements of charged pion asymmetries are an important process in the 
    $\Delta$g global analysis that aims to disentangle all of the partonic contributions to the proton spin.   
    In proton-proton collisions at PHENIX central rapidities, pion production proceed from quark-gluon and 
    gluon-gluon initiated sub processes at the measured $p_{T}$ range. This along with other pion meson properties 
    such as zero spin and pseudo scalar under parity transformations, 
    makes the pion an accessible channel whose importance may elucidate 
    information on the gluon's contribution to the proton's spin.
\section{$\pi$ Mesons}
    Pions, being an isospin triplet, make combining asymmetry measurements from all 
    three pion species for 5 $<$ $p_{T}$ $<$ 12 GeV/c  particularly sensitivity to the sign of $\Delta$g as qg interactions
    dominate pion production in this $p_{T}$ range. 
    Preferential fragmentation of up quarks to positive pions and down quarks to 
    negative pions leads to dominance of ug and dg contributions in the sum over 
    flavors in a factorized pQCD calculation of pion production. In addition, the polarized 
    parton distribution functions(pdf's) are well known to be $\Delta$u $>$0 and $\Delta$d $<$ 0 
    from polarized deep inelastic scattering (DIS) experiments. This dominance of u or d combined with 
    the different signs on their polarized distributions translates into potentially measurable 
    differences in the asymmetries for the different pion species that depend on the sign of $\Delta$g. 
\section{DETECTOR SETUP}
    The PHENIX detector at RHIC has fine-grained calorimetry 100 times finer than 
    previous collider detectors, making particle identification excellent, the resolution of the electromagnetic calorimeter (EMCal)
is $\delta\eta*\delta\phi= 0.01*0.01$. \cite{PHENIX:detector}. 
    Triggering in the central arms allow us to select high $p_{T}$ photons, electrons 
    and charged pions. We select the charged pion signal by requiring a 
    deposition of an energy cluster associated with a charged track in coincidence with the collision trigger(minimum bias trigger). 
    Due to the hadronic response in the EMCal, less than $\frac{2}{3}$ of the charged pions trigger an event\cite{Jia:2003}. 
    We furthermore identify charged pions by selecting particles that produced \v{C}erenkov light in our 
    ring imaging \v{C}erenkov detector(RICH) but did not pass an electromagnetic shower-shape cut.
\section{Signal and Background}

    The measurement presented consists of 0.89 billion events analyzed within a $p_{T}$ range of  5-10 GeV/c, 
    corresponding to an analyzed sample of aproximate 2.3 pb$^{-1}$ , the average polarization was 47\%. 
    The primary source of potential background in this analysis comes from low energy electrons with misreconstructed momentum, and charged hadron tracks firing the RICH. Below 15 GeV/c the only particles that can produce light in the RICH(CO2 radiator) are: 
    electrons(0.017 GeV/c), muons(3.5 GeV/c) and  charged pions(4.7 GeV/c). We performed a series of detector cuts which include: 
    EMCal shower-shape cut of prob $<$ 0. 2  designed to eliminate 80$\%$ of all  electrons, 
    a $p_{T}$-dependent energy cut to remove tracks with mis-reconstructed momentum 
    and a cut of energy/momentum $<$  0.9 to remove the remaining electrons with nearly 
    the correct reconstructed momentum. Muons are not considered a primary source of background for this analysis 
    since primary muon to pion ratios have been previously measured to be less than  $10^{-3}$ at PHENIX central rapidities.
    To estimate the remaining background fraction, the region below 4 GeV/c was fit to a power law.
    Extrapolating the fit under the signal region (solid line, turn on curve times powerlaw) the background 
    is  estimated to be $<$ 5\%.[Figure 1]
   \begin{figure}
 \includegraphics[height=.25\textheight]{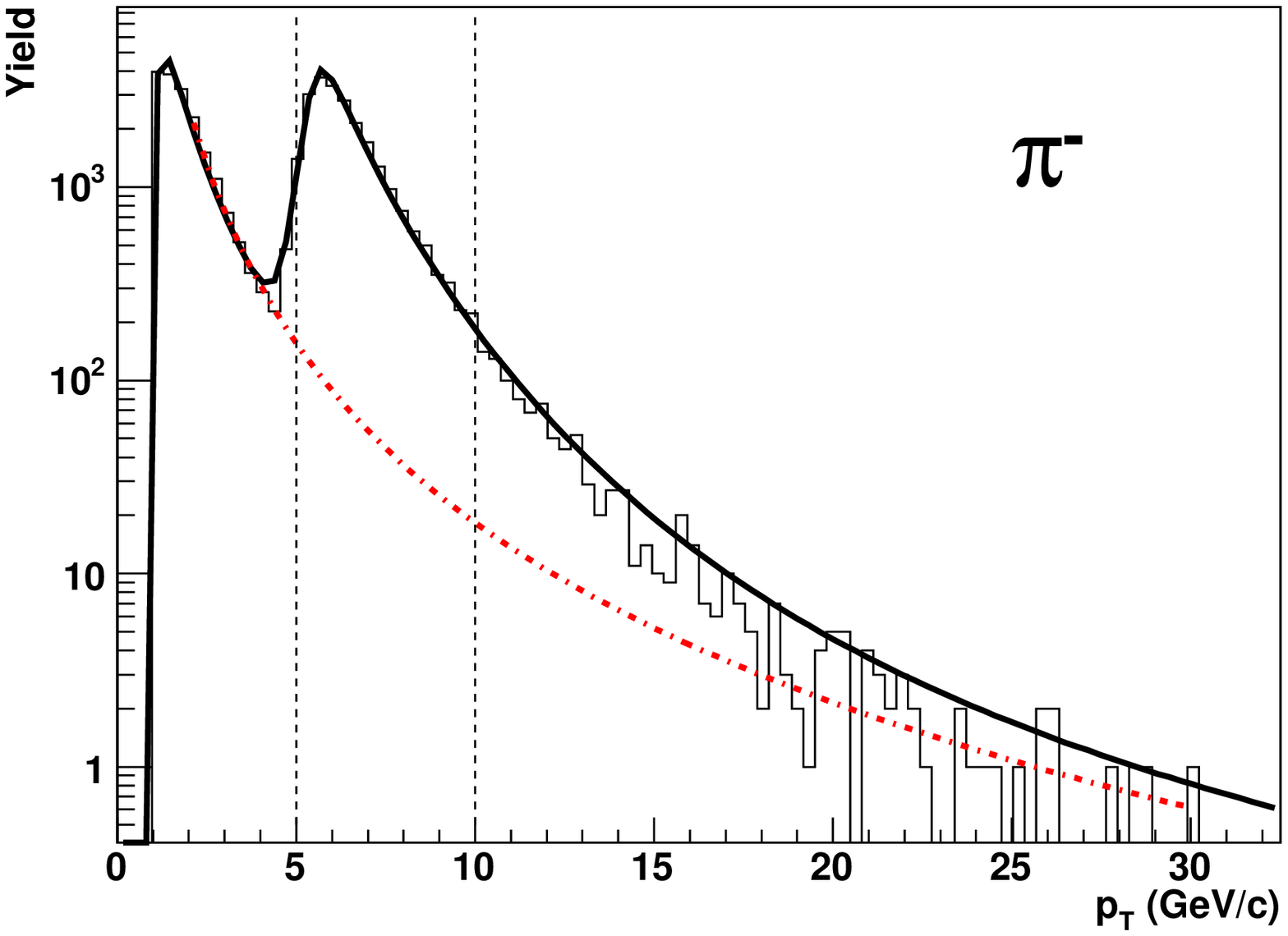}\includegraphics[height=.25\textheight]{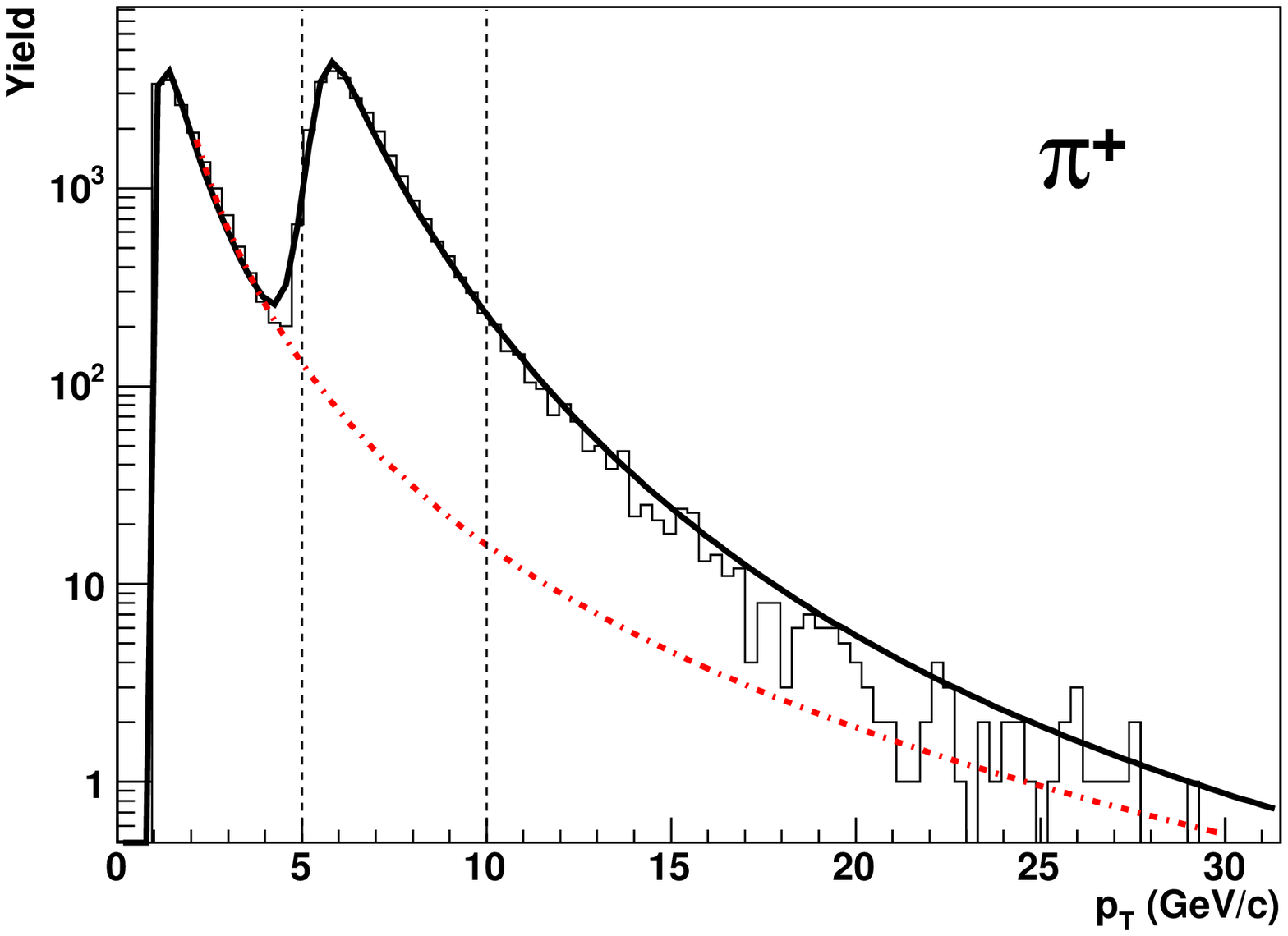}
 \caption{pT Spectrum of negative(left) and positive(right)charged pions fitted to a power law (dashed line)and functional form(solid line)}
 \end{figure}

The $A_{LL}$ and $\delta_{A_{LL}}$ formulae used in this analysis are:
\begin{equation}
 A_{LL} = \frac{\sigma_{++} - \sigma_{+-}}{\sigma_{++} + \sigma_{+-}} = \frac{1}{|P_{Y}||P_{B}|}\frac{N_{++}-RN_{+-}}{N_{++} +RN_{+-}},    
 R =\frac{L_{++}}{L_{+-}}  
\end{equation}
\begin{equation}
 \delta_{A_{LL}} = \frac{1}{|P_{1}||P_{2}|}\frac{2RN_{++}N_{+-}}{(N_{++}+RN_{+-})^2}\sqrt{(\frac{\Delta{N_{++}}}{N_{++}})^2+(\frac{\Delta{N_ {+-}}}{N_{+-}})^2}
\end{equation}
Where $\sigma_{++}(\sigma_{+-})$ is the cross section with the beam in same(opposite) 
helicity configuration, $P_{1}(P_{2})$ is defined as the polarization on separate beams and $N_{++}( N_{+-})$
is the particle yields with the beam in same(opposite)helicity configuration and L is the integrated luminosity.   
\section{Results}
 The measurements for Run-05 double helicity asymmetries for positive and negative charged pions 
 and the associated statistical errors are presented at figure 2.
 A scale uncertainty due to 20$\%$ uncertainty on each beam due to the polarization is not included. The statistical uncertainty in $A_{LL}$ using Run-06 data currently in production is expected to be aproximately 2.7 times smaller.   
 \begin{figure}
 \includegraphics[height=.25\textheight]{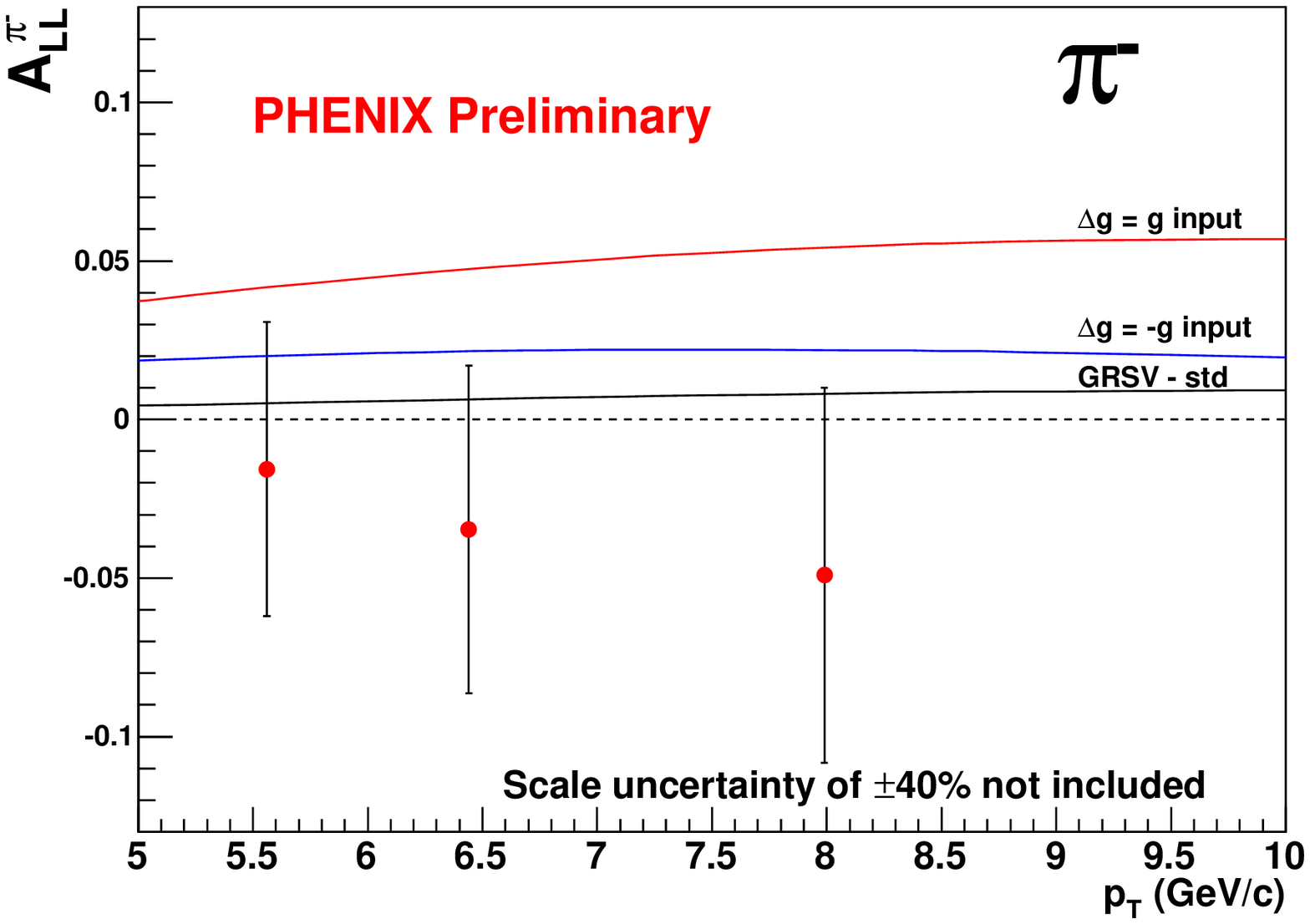}\includegraphics[height=.25\textheight]{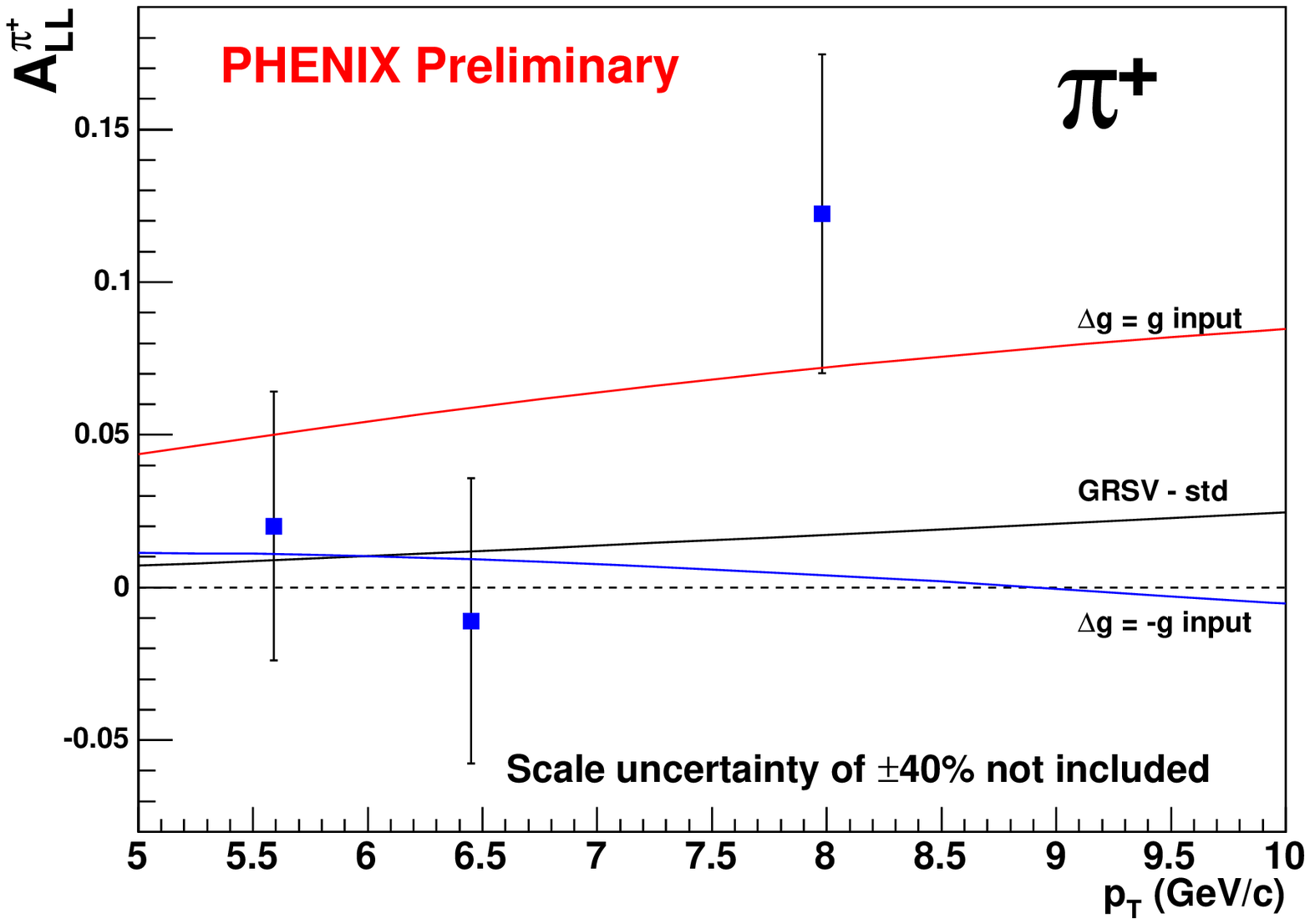}
 \caption{Measured negative(left) and positive(right) charged pion $A_{LL}$. Theory curves are GRSV max\cite{pi:GRSV}($ \Delta$g = g, red), GRSV std(black), GRSV min($\Delta$g = -g, blue).}
\end{figure}

\section{Conclusions}
We presented a first measurement of  $A_{LL}$ at the $p_{T}$ range 5-10 GeV. While the background in this measurement is significantly lower than $\pi^{0}$ \cite{pi0:2003}, for example, current statistical uncertanties do not constrain $\Delta$g. The measurement, in particular the negative(although statistically limited)asymetry observed in negative charged pions, begins to hint to a possible problem in the fragmentation functions used in the parametrizations[Figure 2]. This measurement will be repeated with run-06 data as it becomes available. We expect that this analysis with the higher statistics and a cross-section measurement will be forthcoming and essential in pQCD interpretations and, ultimately, we expect to make a contribution to the global analysis that will determine $\Delta$g.



\bibliographystyle{aipprocl} 

\bibliography{spin2006}

\IfFileExists{SPin2006Proceedings.bbl}{}

\end{document}